\documentclass[twocolumn,showpacs,preprintnumbers,amsmath,amssymb]{revtex4}

\usepackage{graphicx}
\usepackage{dcolumn}
\usepackage{bm}

\begin{document}


\title{Spin-waves in antiferromagnetic single crystal LiFePO$_4$}

\author{Jiying Li\footnote{electronic mail: jiy@ameslab.gov}, Vasile O. Garlea, Jerel L. Zarestky,
and David Vaknin\footnote{electronic mail: vaknin@ameslab.gov}}
\affiliation{Ames Laboratory and Department of Physics and Astronomy, Iowa
State University, Ames, IA 50011
}%


\date{\today}

\begin{abstract}
Spin-wave dispersions in the antiferromagnetic state of single crystal
LiFePO$_4$ were determined by inelastic neutron scattering measurements.  The
dispersion curves measured from the (010) reflection along both {\it a}$^\ast$
and {\it b}$^\ast$ reciprocal-space directions reflect the anisotropic coupling
of the layered Fe$^{2+}$ (S = 2) spin-system.  The spin-wave dispersion curves
were theoretically modeled using linear spin-wave theory by including in the
spin-Hamiltonian in-plane nearest- and next-nearest-neighbor interactions ({\it
J}$_1$ and {\it J}$_2$), inter-plane nearest-neighbor interactions ({\it
J}$_\bot$) and a single-ion anisotropy ({\it D}).   A weak (010) magnetic peak
was observed in elastic neutron scattering studies of the same crystal
indicating that the ground state of the staggered iron moments is not along
(010) direction, as previously reported from polycrystalline samples studies,
but slightly rotated away from this axis.
\end{abstract}

\maketitle

\section{Introduction }

\begin{figure} [htl]
\centering
\includegraphics[width = 0.43\textwidth] {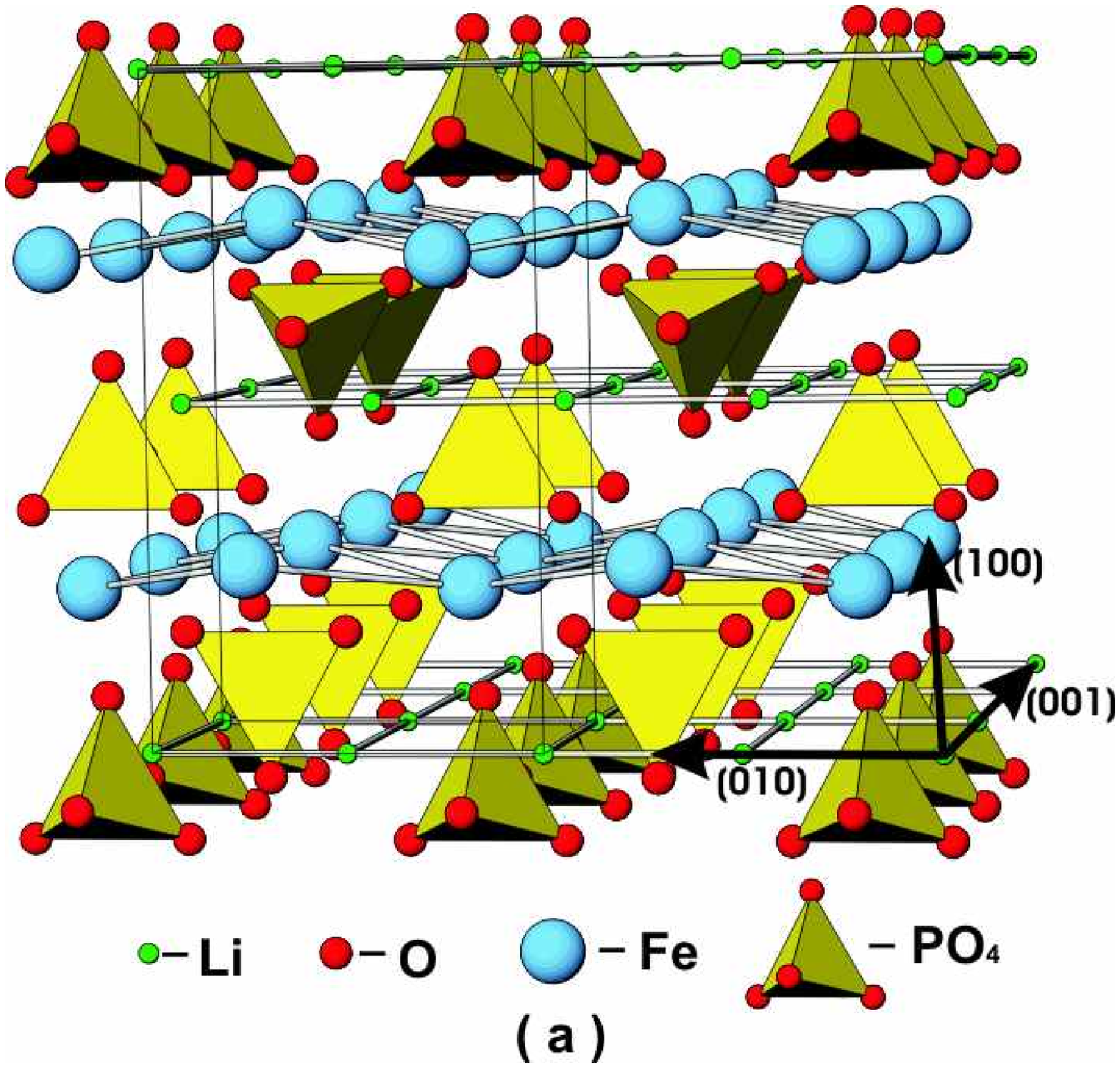}
\includegraphics[width = 0.43\textwidth] {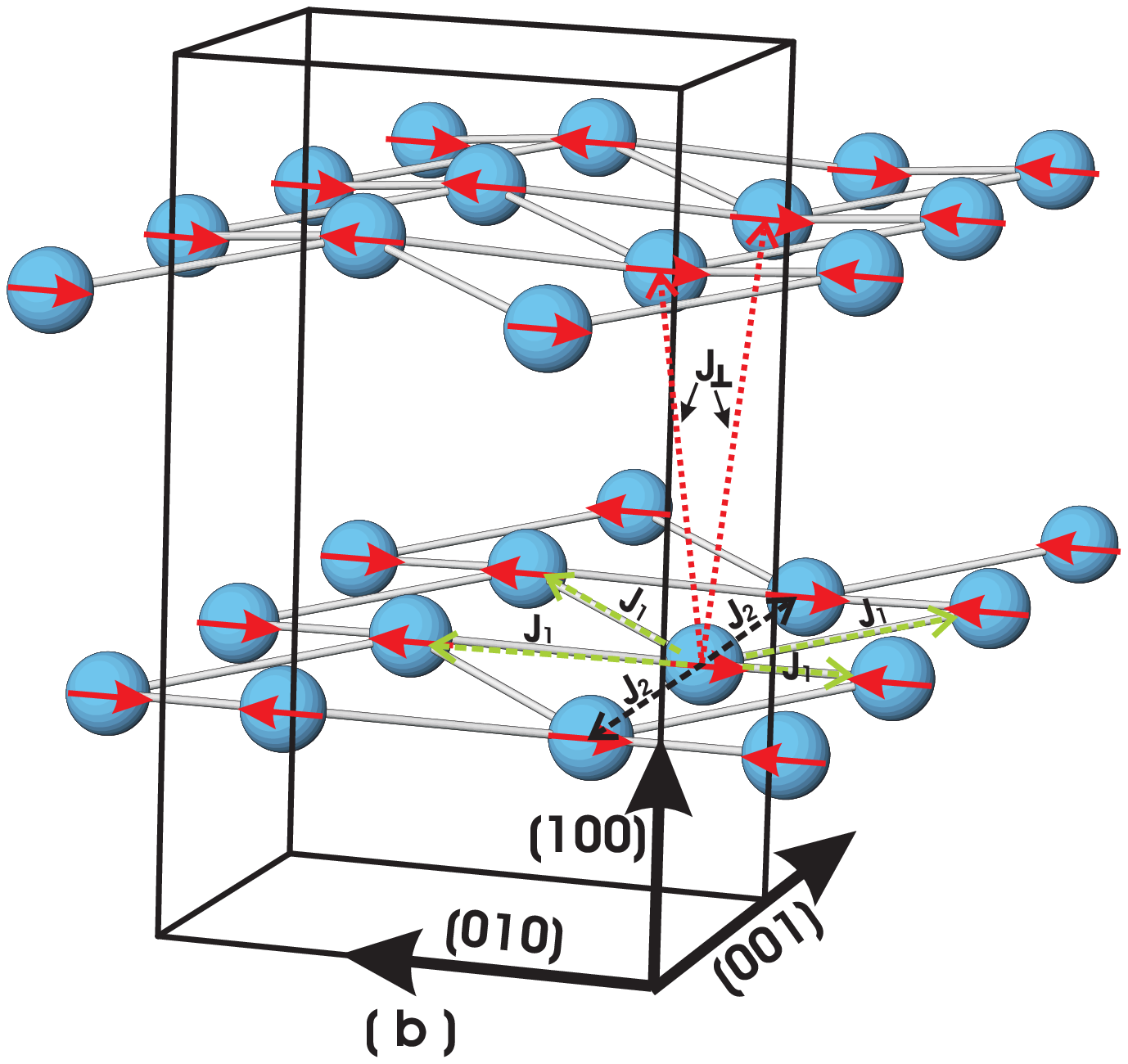}
\caption{\label{Str}(color online) (a) Atomic structure of LiFePO$_4$. The
Fe$^{2+}$ ions form buckled layers stacked perpendicular to the [100]
crystallographic direction. The ground state of LiFePO$_4$ is collinear
antiferromagnetic with the average moment along {\it b} direction. (b) Spin
arrangement of the two Fe$^{2+}$ layers, the in-plane nearest and next-nearest
neighbor interactions {\it J}$_1$ and {\it J}$_2$ and inter-plane nearest
neighbor inaction {\it J}$_\bot$ are labeled.}
\end{figure}

Lithium-orthophosphates Li{\it M}PO$_4$ ({\it M} = Mn, Fe, Co, Ni) have
attracted a renewed interest in recent years, both for their relatively high
lithium ionic-conductivity that can potentially be applied in rechargeable
battery technology \cite {Tarascon01}, and for their intriguing magnetic
properties, in particular, the strong magneto-electric (ME) effect they exhibit
\cite {Fiebig05}.  In this regard, of particular importance is LiFePO$_4$, as
it has already been tested as a high-potential cathode in secondary Li-ion
rechargeable battery \cite {Goodef97, Chung02, Herle04, Tarascon05}.  Like
other members of the lithium-orthophosphates, LiFePO$_4$ is an insulator
adopting the {\it Pnma} space group \cite {Geller60, Megaw73}.  In this
structure, the Fe$^{2+}$ ion occupies the center of a slightly distorted {\it
M}O$_6$ octahedron that shares oxygen anions with a PO$_4$ tetrahedron forming
 a closely packed oxygen framework.  The Fe$^{2+}$ ions (S = 2) form corrugated
layers that are stacked along the [100] crystallographic axis, as shown in Fig.
\ \ref{Str}(a).  Nearest neighbors in the {\it b-c} plane are coupled
magnetically by a relatively strong exchange interaction, $J_1$ through an
Fe-O-Fe oxygen-bond, whereas in-plane next-nearest-neighbors are coupled
($J_2$) via Fe-O-O-Fe \cite {Dai05} (see Fig. \ \ref{Str}(b) for definition of
exchange couplings). Interlayer magnetic coupling is mediated by a phosphate
ion via an Fe-O-P-O-Fe bonding \cite {Mays63}. Thus, the olivine family of
Li{\it M}PO$_4$ exhibits highly anisotropic properties which are between those
of two- (2D) and three-dimensional (3D) systems \cite {Vaknin99, Vaknin02}.

The magnetic properties of Li{\it M}PO$_4$ systems have been studied since the
early 1960s \cite {Mercier67, Santoro66, Santoro67}.  They all undergo an
antiferromagnetic transition at low temperatures to a similar magnetic
arrangement differing only in the orientation of the staggered spins.  Nearest-
neighbor (NN) spins in the {\it b-c} plane are anti-parallel and the stacking
along the $a$-axis is such that ferromagnetic sheets perpendicular to the
$b$-axis are formed; nearest neighbor sheets are anti-parallel giving rise to
(010) fundamental magnetic reflection that, depending on the direction of the
magnetic moment, can be intense or extinct.  Earlier neutron diffraction
studies of polycrystalline samples showed that the magnetic space group of
LiFePO$_4$ and LiCoPO$_4$ is {\it Pnma}' with the spins oriented along {\it b}
crystallographic direction (i.e., the (010) reflection is absent), and {\it
Pnm}'${\it a}$ magnetic space group for LiNiPO$_4$ and LiMnPO$_4$ with the
spins aligned parallel to the $c$-axis (i.e., strong intensity at the (010)
reflection) \cite {Vaknin99, Vaknin02, Santoro67, Goni96, Arcon04a,
Kharchenko03}.  Recent neutron diffraction measurements of single crystal
LiCoPO$_4$ reported finite intensity at the (010) reflection, interpreted in
terms of a ground state with a spin direction that is slightly rotated from the
{\it b}-axis \cite {Vaknin02}.  Weak ferromagnetism has been reported for
LiMnPO$_4$ \cite {Kharchenko03} and LiNiPO$_4$ \cite {Arcon04b} at a
temperature below {\it T}$_N$.

Recently, the magnetic structure and properties of lithium orthophosphates have
been reexamined theoretically and experimentally \cite {Streltsov93, Rousse03}.
Rousse et al. \cite {Rousse03} reported neutron diffraction results from
polycrystalline samples confirming the collinear structure below {\it T}$_N$ =
52 K.   Magnetic properties of LiFePO$_4$ investigated by M\"{o}ssbauer
spectroscopy and magnetization measurement determined that {\it T}$_N$  = 50 K
\cite {Zhi04}.   Theoretical estimations of the spin exchange coupling by
spin-dimer analysis, while neglecting the single ion-anisotropy, yield the
following values {\it J}$_1$ = 1.08 meV, {\it J}$_2$ = -0.4 meV and {\it
J}$_\bot$ = -0.92 meV \cite {Dai05}.   However, so far there has been no
experimental determination of the exchange coupling among Fe$^{2+}$ spins and
of the single-ion anisotropy in LiFePO$_4$, for comparison with the theoretical
predictions.  Knowledge of exchange couplings and single-ion-anisotropy is also
important for understanding the origin of the strong magneto-electric (ME)
effect in LiFePO4 \cite {Mercier67}.  All lithium orthophosphates exhibit a
strong yet anomalous linear magnetoelectric (ME) effect with respect to the
observed ME tensor components,  $\alpha$$_{xy}$,  $\alpha$$_{yx}$, for
LiFePO$_4$ and LiCoPO$_4$, and  $\alpha$$_{xz}$,  $\alpha$$_{zx}$ for
LiNiPO$_4$, as expected with their respective antiferromagnetic point groups
mmm' and mm'm \cite {Mercier68,Mercier71,Rivera94,Vaknin04}. In particular, the
ME effect measurements of LiFePO$_4$ as a function of temperature reveal a
decrease of the ME coefficient along one direction $\alpha$$_{yx}$(T) below a
maximum close to {\it T}$_N$ \cite {Mercier68, Mercier71}. Detailed
determination of the magnetic structure using neutron diffraction from single
crystals can shed light on the origin of these anomalies.

Herein, we report measurements of spin-wave dispersion curves of single crystal
LiFePO$_4$ by inelastic neutron scattering measurements.  Spin wave dispersion
curves can provide the values of exchange interactions and the single ion
anisotropy.  The measured dispersion curves were modeled using linear spin wave
theory by including the in-plane nearest-neighbors (NN) and
next-nearest-neighbors (NNN) spin couplings the inter-plane nearest-neighbor
spin coupling and the single-ion anisotropy.  We have also employed
single-crystal elastic neutron diffraction techniques to investigate whether
there are subtle deviations from the previously reported magnetic structure
determined from neutron diffraction measurements of polycrystalline samples.
\section{Experimental Details}
LiFePO$_4$ single crystals were grown by standard flux growth technique (LiCl
was used as the flux) from a stoichiometric mixture of high purity FeCl$_2$
(99.999$\%$ Aldrich) and Li$_3$PO$_4$ (99.999$\%$ Aldrich) \cite {Fomin02}. The
grown single crystals have a dark greenish color. The composition and structure
were confirmed by carrying out Rietveld analysis of the X-ray powder
diffraction (XRD) data, using the GSAS software package \cite {Larson90}.  No
extra peaks from impurities were detected in the XRD pattern.  Powder, for the
XRD, was produced by crushing typical isolated single crystals from the melt.
The lattice parameters yielded from the refinement ($a = 10.337 $ {\AA}, $b =
6.011 $ {\AA}, and $c = 4.695 $ {\AA}) are in good agreement with literature
values \cite {Santoro67, Streltsov93, Rousse03}.

Neutron scattering measurements were carried out on the HB1A triple axis
spectrometer at High Flux Isotope Reactor at Oak Ridge National Laboratory.  A
monochromatic neutron beam of wavelength $\lambda $ = 2.366 \AA\ (14.6125 meV,
$k_{o}=2\pi /\lambda =2.656$\AA$^{-1}$) was selected by a double monochromator
system, using the (0 0 2) Bragg reflection of highly oriented pyrolytic
graphite (HOPG) crystals.  HOPG crystal was also used as analyzer for both the
elastic and the inelastic studies.  The $\lambda $/2 component in the beam was
removed by two HOPG filters located before and after the second monochromator.
The collimating configuration 40$^\prime $- 40$^\prime $ - Sample - 34 $^\prime
$- 68 $^\prime $ was used throughout the experiments.  Temperature measurements
and control were achieved by a conducts LTC-20 using Lake Shore silicon-diode
temperature sensors.  An irregular shaped single crystal (weight $\sim$ 0.4 g)
was folded in aluminum foil and mounted on a thin aluminum post. It was then
sealed in an aluminum can under helium atmosphere and loaded onto the tip of a
closed-cycle helium refrigerator (Displex).  Two temperature sensors were
mounted on the cold-tip of the Displex and on the sample can.  The temperature
was controlled using the cold-tip sensor.  The temperature difference between
the two sensors was about 0.2 K over the temperature range investigated.

\section{Results and Discussion \protect}

\subsection{Elastic neutron scattering}
\begin{figure}[htl]
\centering
\includegraphics[width = 0.43\textwidth] {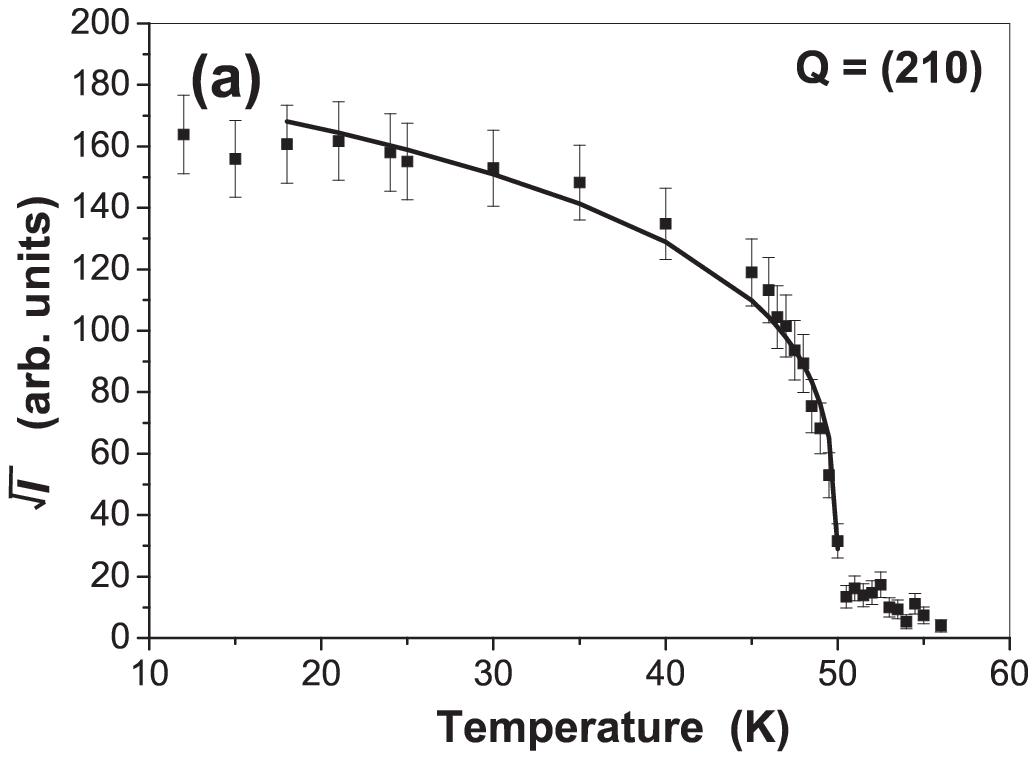}
\includegraphics[width = 0.43\textwidth] {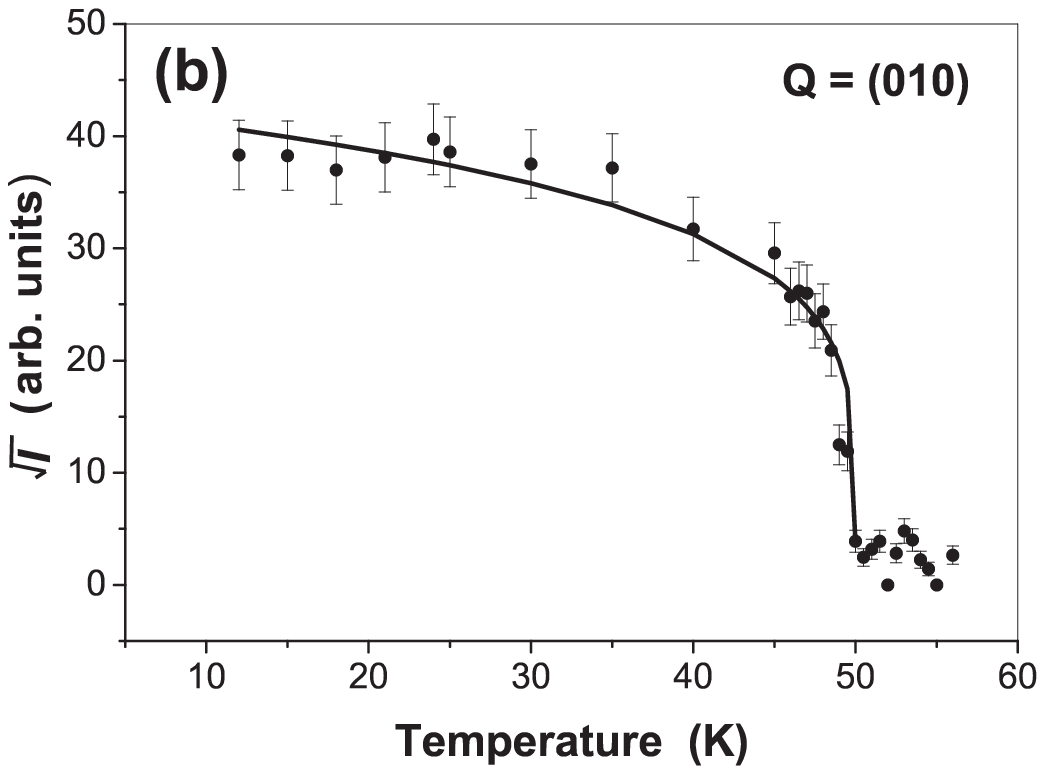}
\caption{Temperature dependence of the square root of the normalized integrated
intensity at two reflection peaks (210) (a) and (010) (b). The transition
temperature obtained from the fitting is {\it T}$_N$ = 50 $\pm$ 0.5K and the
critical exponent $\beta$ = 0.27 $\pm$ 0.03.} \label{Order}
\end{figure}

The LiFePO$_4$ crystal was oriented such that the {\it a-b} plane coincided
with the horizontal scattering plane of the spectrometer.  The elastic
measurements confirmed that the magnetic structure of LiFePO$_4$ is
antiferromagnetic with the main direction of the moment oriented along the {\it
b}-axis.  However, contrary to the previous neutron diffraction experiments
performed on powder samples, we have detected the presence of the (0 1 0)
reflection. The intensity of this peak is relatively weak but its intensity
follows a similar temperature dependence as that of a stronger magnetic peak
(210) (see Fig.\ \ref{Order}).  It is worth noting that the (0 1 0) peak is
forbidden by the symmetry of the {\it Pnma}$^\prime$ magnetic space group
previously proposed to describe the spin arrangement in LiFePO$_4$ \cite
{Santoro67, Streltsov93}. In the {\it Pmna} crystal symmetry, the point group
of the Fe 4{\it c} site is {\it m}$_y$ and the only allowable magnetic point
groups are, therefore, {\it m}$_y$ with the Fe magnetic moments along the
$b$-axis (perpendicular to the mirror (010) plane) and {\it m}$^\prime$$_y$
with the magnetic moment lying in the mirror (010) plane.  The magnetic
contribution to the (010) peak indicates that the ordered moment is not
strictly oriented along the $b$-axis and a small component perpendicular to
this axis is present.  This implies a lowering of the symmetry of the magnetic
space group where both magnetic components (along and perpendicular to
$b$-axis) are allowed.  From the intensity ratio of the two reflections {\it
I}(010)/ {\it I}(210) at low temperatures, we can estimate the angle of the
staggered moment with respect to the {\it b}-axis, by using the following
relation,
\begin{equation}
\frac{F_{(010)}\sin(\alpha_{(010)})}{F_{(210)}\sin(\alpha
_{(210)})}=\sqrt{\frac{I_{(010)}}{I_{(210)}}\frac{\sin(2\theta_{(210)})}
{\sin(2\theta_{(010)})}}\frac{f_{(210)}}{f_{(010)}} \label{Eq1}
\end{equation}
where, {\it F}$_{(210)}$ and {\it F}$_{(010)}$ are the magnetic structure
factors of peaks (210) and (010),  $\alpha_{ (210)}, \alpha_{(010)}$ are the
angles between the scattering vector of reflections (210), (010) and the
magnetic moment, and {\it f}$_i$ are the corresponding form
factors\cite{Vaknin02}. Using {\it f}$_{(210)}$/{\it f}$_{(010)}$$ \approx$
0.85, we estimate the moments are rotated by 7.5 $\pm$ 0.5 deg toward {\it
c}-axis or 3 $\pm$ 0.5 deg toward {\it a}-axis.
The ratio between the magnetic and nuclear contributions to the peak
intensities of reflections ({\it I}$_{mag}$/{\it I}$_{nuc}$) can be used to
determine the average magnetic moment, $\mu$ from
\begin{equation}
\mu =
\sqrt{\frac{I_{mag}}{I_{nuc}}\frac{|F_{nuc}|^2}{|F_{mag}|^2}\frac{1}{f^2({Q})\sin^2\alpha}}
\label{Eq2}
\end{equation}
where for the reflection in question, $F_{nuc}$, and $F_{mag}$ are the nuclear
and magnetic structure factors, $I_{nuc}$ and $I_{mag}$ are the nuclear and
magnetic intensities and $f(Q)$ is the magnetic form factor of Fe$^{2+}$ at
momentum transfer $Q = 2k_0\sin\theta$. $I_{mag}$ can be calculated from the
peak intensity difference at temperatures above and below $T_N$. Using the peak
intensities of (210) at 300 K and 10 K to calculate the $I_{mag}$ and
$I_{nuc}$, and using $f_{(210)}$  = 0.85, the calculated average magnetic
moment $\mu$ for Fe is 3.93 $\pm $ 0.05 $\mu_B$, which is very close to the
values of 3.99 $\mu_B$ and 3.8 $\mu_B$ obtained in Refs. \cite {Xu04} and \cite
{Yamada01}.

To determine the temperature dependence of the order parameter, the (010) and
(210) reflections were monitored as a function of temperature in the range 10
to 60 K.  Figure\ \ref{Order}(a) and (b) show the square root of the integrated
intensity ($\sqrt{I}$) {\it versus} temperature.  The $\sqrt{I}$ quantity is
proportional to the antiferromagnetic staggered magnetization, i.e., the order
parameter. It was fitted to the following power law function near the
transition temperature,
\begin{equation}
\sqrt{I} \mbox{ }{\propto}\mbox{ } M^{\dagger} = M^{\dagger}_0t^{\beta}
\label{eq3}
\end{equation}
where, {\it M}$^\dag_0$ is the sub-lattice magnetization at {\it T} = 0 K, {\it
t} = (1-{\it T}/{\it T}$_N$) is the reduced temperature, and $\beta$ is the
critical exponent.  For the two magnetic peaks (010) and (210) the obtained
transition temperatures are the same, {\it T}$_N$ = 50 $\pm$ 0.5 K and the
critical exponent $\beta$ is 0.27 $\pm$ 0.03.  The transition temperature is
very close to the values reported in the literature 50 $\pm$ 2 K \cite
{Santoro67, Streltsov93, Zhi04}. The critical exponent $\beta$ is slightly
smaller than that calculated for the 3D Ising model ($\beta$= 0.32) \cite
{Collins89}.  The temperature dependent background-like scattering above {\it
T}$_N$ and below $\sim$ 60 K indicates some kind of critical scattering due to
short-range order formation or due to a dimensionality cross-over.

\subsection{Inelastic neutron scattering}
\begin{figure}[htl]
\centering
\includegraphics[width = 0.43\textwidth] {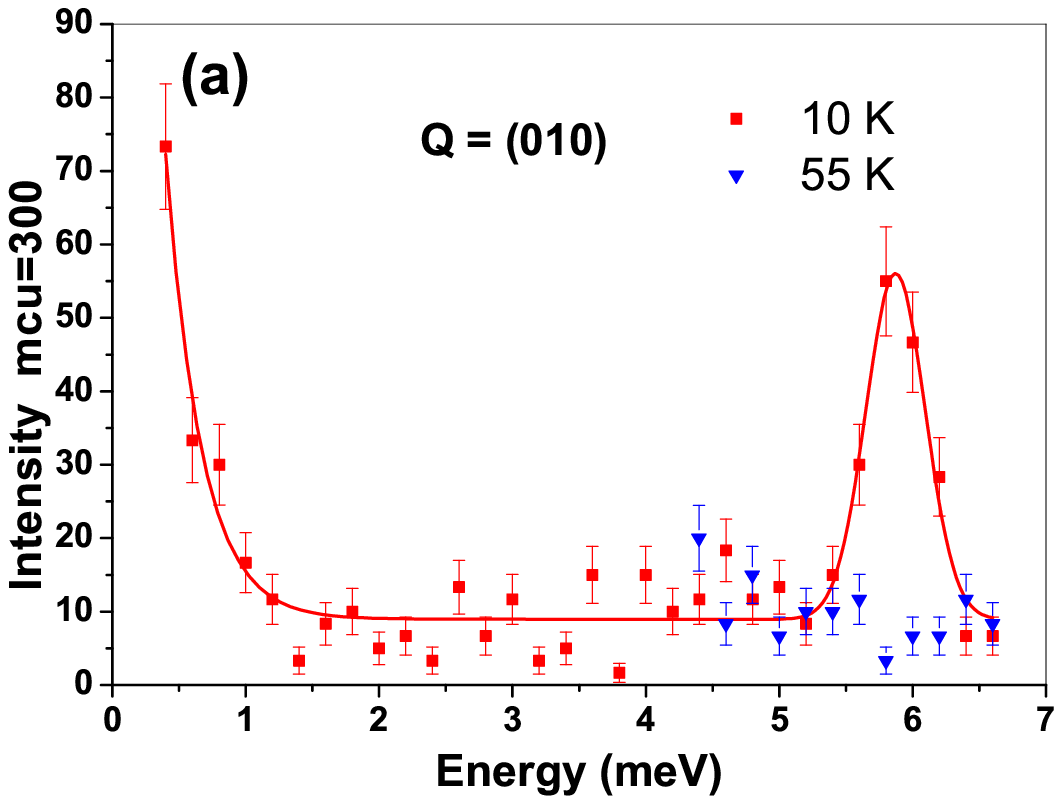}
\includegraphics[width = 0.43\textwidth] {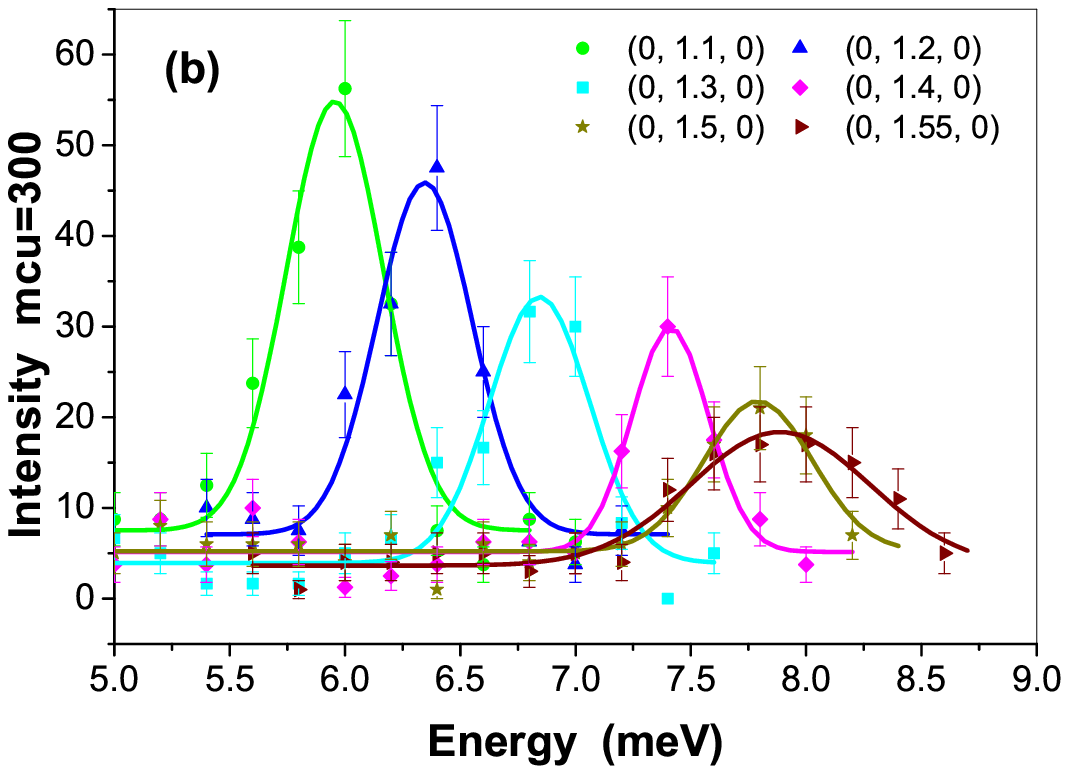}
\includegraphics[width = 0.43\textwidth] {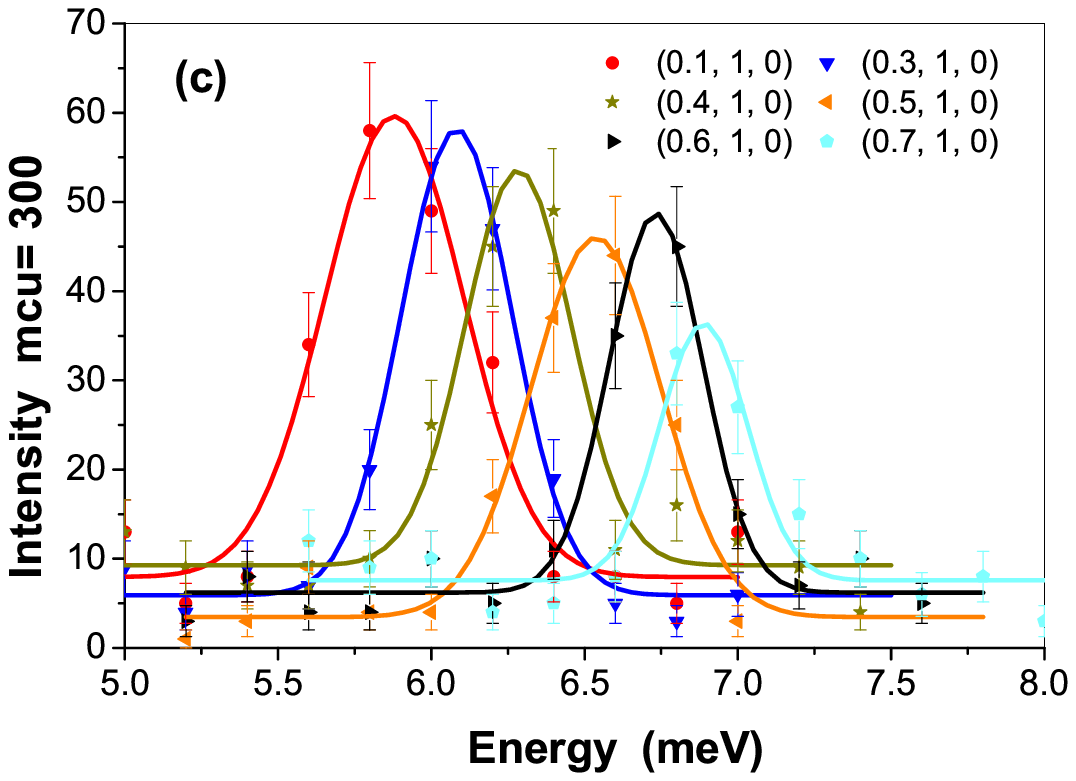}
\caption{(color online) (a) Neutron scattering intensity as a function of the
energy transfer {\it E} at (010) peak at 10 K and 55 K. (b) Constant-Q scans
taken at 10K, at different wave-vectors (0, 1+ $\xi$, 0) and (c) Constant-Q
scans at ($\xi$, 1, 0).}
\label{Inelastic}
\end{figure}

The spin wave excitations were measured at 10 K along the ($\xi$, 1, 0) and (0,
1+$\xi$, 0) reciprocal space directions, for energy transfers (energy loss
mode) ranging from 1 to 8.5 meV.  As illustrated in Fig. \ref{Inelastic}
(a,b,c), well defined dispersive magnetic-modes of resolution-limited
energy-width were observed at all wave vectors. A typical constant-q scan,
performed at the zone center (010), is shown in Fig. \ref{Inelastic} (a),
indicating a single excitation at an energy transfer of 5.86 $\pm$ 0.04 meV.
The increase in intensity at energy transfer below approximately 1 meV is due
to the quasielastic scattering from the newly observed (0 1 0) magnetic Bragg
peak. Above the transition temperature ($T_N$ = 50 K), at approximately 55 K,
the inelastic peak at 5.86 meV disappears, confirming its magnetic origin. Such
an energy gap in the dispersion curve is usually driven by single ion
anisotropy. A similar energy gap of 2 meV at 2 K, was also observed in the
LiNiPO$_4$, and it was found to decrease with increasing the temperature \cite
{JLi05}.  In the case of LiFePO$_4$, measurements performed at different
temperatures indicate that the energy gap is temperature independent.  The
inelastic scattering signal measured at different constant wave-vectors $\xi$
along the (100) and (010) reciprocal-space directions, at 10 K, are shown in
Figure \ref{Inelastic} (b, c).  The data were fitted to Gaussian functions
(solid line in Fig. \ref{Inelastic} (a,b, c)) where the background was assumed
to be constant.

\begin{figure} [htl]
\centering
\includegraphics[width = 0.43\textwidth] {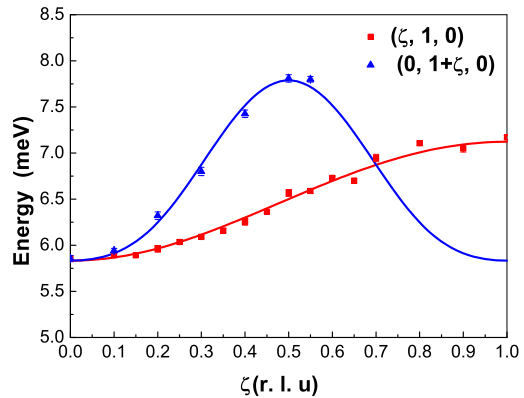}
\caption{\label{dispersion}(color online) Spin-wave dispersion curves along the
{\it b}$^*$ and {\it a}$^*$ reciprocal space  directions. Solid lines are fits
obtained from Linear Spin-wave Theory using Eq(5).}
\end{figure}

The spin-wave dispersion branches deduced from these fits, for both {\it b}$^*$
and {\it a}$^*$ reciprocal-space directions, are plotted in Figure
\ref{dispersion}.  It is shown that the dispersion curves montonically increase
in energy with $\xi$, and that the spin-waves propagating in the plane along
the (010) direction have higher frequencies than those propagating
transversally along the (100).  Qualitatively, this behavior reflects the
anisotropy in the strength of exchange couplings in the system;  as expected
the in-plane exchange couplings are much stronger than those between planes.

To construct a Hamiltonian for the spin system, we recall that in LiMPO$_4$
olivine family the in-plane super-exchange or super-super exchange interactions
between nearest and next-nearest neighboring Fe$^{2+}$ ions ({\it J}$_1$ and
{\it J}$_2$) are expected to be much stronger than that between the nearest
inter-plane neighbors ({\it J}$_\bot$) \cite {Mays63, Shi05}. Therefore, we
propose the following Hamiltonian,
\begin{eqnarray}
  {\cal H} = -J_1\sum_{i, \delta}(S_iS_{i+\delta})-J_2\sum_{i, \xi}(S_iS_{i+\xi})\nonumber\\
-J_{\bot}\sum_{i,\delta_\bot} (S_iS_{\delta_\bot} )+D\sum_i(S^z)^2
\end{eqnarray}
where, $J_1$ and $J_2$ are the in-plane NN and NNN coupling constants,
respectively,  and $J_\bot$ is the inter-plane NN coupling constant. The
illustrations of {\it J}$_1$, {\it J}$_2$ and $J_\bot$ are shown in Figure\
\ref{Str}(b).  {\it D} is the single-ion anisotropy constant quantifying the
tendency of the spins to align along the easy axis (the $S^z$ component is
defined to be along the direction of the moment at the ground state - {\it
b}*).  The Ising like ground state of the system is believed to be invoked by
the single-ion anisotropy term which comes about from crystal field effects and
spin-orbit coupling \cite {Jongh90}. Using the antiferromagnetic spin-wave
theory\cite {Anderson52, Kubo52}, the lattice with {\it N} sites was divided
into two sublattices A and B, where  nearest-neighbors of an Fe$^{2+}$ site in
one sub-lattice are all sites in the other sublattice. The next-nearest
neighbors of an Fe$^{2+}$ site are in the same sublattice. The magnon
dispersion curves were calculated using the Holstein-Primakoff spin operator
transformation to linear approximation (i.e., linear spin-wave theory \cite
{Squires78}).  The resulting spin-wave dispersion is given by
\begin{equation}\label{Eq5}
\hbar\omega = \sqrt{A^2 - F^2}
\end{equation}
where {\it A} = (2{\it J}$_1${\it ZS} - 2{\it J}$_2${\it ZS} - 2{\it
J}$_\bot${\it ZS} + 2{\it J}$_2${\it ZS}$\gamma$$_{3N}$ + 2{\it J}$_\bot${\it
ZS}$\gamma$$_{\bot}$ + 2{\it DS}) and {\it F} = 2{\it J}$_1${\it
ZS}$\gamma$$_{2N}$, in which Z is the number of the nearest neighbors Z = 4, S
=2 for Fe$^{2+}$. $\gamma$$_{2N}$, $\gamma$$_{3N}$ and $\gamma$$_{\bot}$ are
calculated using the following equation:
\begin{equation}
\gamma_{(2N, 3N, \bot)} = {\frac {1}{Z}}\sum_i{e^{i{\bf Q \cdot r}}}
\end{equation}
where {\it\bf r} = ($\delta, \xi, \delta_\bot$) are the components of vectors
to the intra-plane nearest, next-nearest neighbors and to the inter-plane
nearest-neighbor. We get
\begin{equation}
\gamma_{2N} = \cos(\pi k_y)\cos(\pi k_z)
\end{equation}
\begin{equation}
\gamma_{3N} = \frac{1}{2}(\cos(2\pi k_y) + \cos(2\pi k_z))
\end{equation}
\begin{equation}
\gamma_{\bot} = \cos(\pi k_x)\cos(\pi k_y)
\end{equation}
The experimental data along the (0, 1+$\delta$, 0) and ($\delta$, 1, 0)
directions in Fig. \ref{dispersion} were fitted to Eq. ~(\ref{Eq5}) yielding
the following values: $J_1$ = -0.662 $\pm$ 0.02 meV, $J_2$ = -0.27 $\pm$ 0.02
meV, $J_\bot$ = 0.021 $\pm$ 0.001 meV and $D$ = -0.37 $\pm$ 0.01 meV.

The in-plane nearest- and next-nearest-neighbor coupling constants
quantitatively agree with theoretical calculations\cite{Dai05}, $J_1$ = -1.08
meV and $J_2$ = -0.4 meV \cite {Dai05}.  The two spin couplings, $J_1 < 0$ and
$J_2 < 0$, compete oppositely over the alignment of in-plane NNN spins; whereas
$J_1$ leads to parallel alignment of NNN spins $J_2$ favors their antiparallel
alignment. Such competing interactions can lead to incommensurate
phases\cite{Bak82} which were not found in this system. However, incommensurate
phases have been reported for the isostructural LiNiPO$_4$\cite{Vaknin04}. The
inter-plane coupling $J_\bot$ = 0.021 meV determined in this study is
significantly smaller than the theoretical one $J_\bot$ = -0.92 meV
\cite{Dai05}.  It should be noted that single-ion anisotropy was not considered
in the theoretical calculations \cite {Dai05}.

To summarize, we have measured spin-wave dispersions and determined spin
exchange couplings in LiFePO$_4$.  Our results show that although there are
competing interactions between NN and NNN spins in LiFePO$_4$, they do not lead
to more complicated, incommensurate or non-colinear, magnetic structures. This
is in contrast to the observation of incommensurate magnetic phases in
LiNiPO$_4$\cite{Vaknin04}.  These competing interactions may explain the
observation of weak-ferromagnetism in Li$M$PO$_4$ systems\cite{Kharchenko03}.
They may also be related to the observation, in this study, that the staggered
magnetic moment is not aligned along a principal direction. From the gap in the
spin wave dispersion curve, we have been able to extract the single-ion
anisotropy in LiFePO$_4$ using linear spin-wave theory.

\begin{acknowledgments}
We thank R. J. McQueeney, S. Chang, and J. Q. Yan at Ames Laboratory for the
helpful discussions. This work was supported (in part) under the auspices of
the United States Department of Energy. The HFIR Center for Neutron Scattering
is a national user facility funded by the United States Department of Energy,
Office of Basic Energy Sciences- Materials Science, under Contract No.
DE-AC05-00OR22725 with UT-Battelle, LLC.  The work was supported by the
Department of Energy, Office of Basic Energy Sciences under contract number
W-7405-Eng-82.
\end{acknowledgments}
\newpage 

\end{document}